# Author details

Snezna Bizilj Schmidt
School of Information Technology & Systems
Faculty of Science & Technology, University of Canberra, Bruce, Australian Capital Territory, Australia.
snezna.schmidt@canberra.edu.au (corresponding author)
ORCID ID: https://orcid.org/0009-0004-2415-1206

Nathan M. D'Cunha, PhD (Health), BHumNutr (Hons)
Associate Professor, Human Nutrition, School of Exercise and Rehabilitation Sciences and
Theme Lead (Dementia and Cognition), Center for Ageing Research and Translation
Faculty of Health, University of Canberra, Bruce, Australian Capital Territory, Australia.
ORCID ID: https://orcid.org/0000-0002-4616-9931

Stephen Isbel, HScD, MOT, MHA, BAppSc (OT), GCTE
Professor, Occupational Therapy, School of Exercise and Rehabilitation Sciences and
Theme Lead (Innovative Care Models), Center for Ageing Research and Translation, Faculty of Health, University of Canberra, Bruce, Australian Capital Territory, Australia.
ORCID ID: https://orcid.org/0000-0001-5355-3205

Blooma John, PhD
Associate Professor and Capability Lead in the School of Information Technology & Systems
Faculty of Science & Technology, University of Canberra, Bruce, Australian Capital Territory, Australia.
ORCID ID: https://orcid.org/0000-0001-6725-6025

Ram Subramanian, PhD
Associate Professor in the School of Information Technology & Systems
Faculty of Science & Technology, University of Canberra, Bruce, Australian Capital Territory, Australia.
ORCID ID: https://orcid.org/0000-0001-9441-7074

# Design principles to Increase Technology Self-efficacy for Older Australians with Mild Cognitive Impairment (MCI) and Older Carers

## Design principles to maximise technology self-efficacy for older adults with MCI

The number of people with age-related physical or cognitive impairments is increasing due to the world's ageing population. Technology has the potential to support independent living, and to achieve aged and health service efficiencies; however, there are gaps in our understanding of factors that motivate technology adoption and ongoing use by older adults, especially those with cognitive impairments. This study aims to explore motivators and enablers for technology adoption and ongoing use by older Australians with mild cognitive impairment (MCI) and their carers, to identify technology design principles and guidelines that maximise adoption. Semi-structured interviews were used to gather data about individual demographics, needs, priorities, lifestyle, challenges, and experiences with technology. Results of inductive, reflective, thematic analysis indicate that a desire for independence, autonomy and quality of life motivate use of technology, and perceived technology self-efficacy and IT literacy are enablers. The Protection Motivation Theory illustrates that constant technology change is a disabler for technology adoption and sustained use, because it lowers perceived technology self-efficacy and IT literacy, and reduces confidence to use technology. Two high-level technology design principles and related guidelines are proposed, grounded in theory and aligned with Bandura's four sources of self-efficacy. These design principles and guidelines are intended to increase feelings of self-efficacy and confident use of technology, while also reducing adverse impacts of technological change and encouraging sustained technology adoption by older adults with MCI to support independent living and quality of life.

### 1 INTRODUCTION & BACKGROUND

Incidence of age-related physical and cognitive limitations is increasing as a result of the world's ageing population [1], which adds pressure to already overloaded aged care and health services. Technology has been shown to reduce resource and financial requirements, and increase healthcare service productivity [2], so attention is being directed on development of a variety of technologies to support healthy ageing and independent living. Mild cognitive impairment (MCI) results in decline greater than expected for a person's age and education level, in one or more cognitive domains or affective attributes such as memory, language, executive functioning, circadian rhythms, or visuospatial ability [3-5]. MCI may affect up to 20% of people over 65, and it is estimated up to 830,000 Australians over 65 may have MCI as of 2024 [6, 7], and this number will increase as Australia's population ages. People with MCI may forget things more frequently, lose items more often, struggle to remember words or have problems with language, have trouble making decisions or following instructions, lose their train of thought, have new difficulties regulating emotions, behave more impulsively, have troubles with visual perception and sleeping, experience motor coordination issues, or be less able to follow daily routines [7]. People experience MCI symptoms uniquely and individually, and a variety of technology solutions are being considered to assist older adults with MCI to live independently [8-11]. Recent studies illustrate technology use by older adults is increasing, however it is still lower than younger people [12-16]. Research has confirmed that older adults with MCI already engage with technology and online services to support a range of needs and support quality of life [17-19], and there is encouragement for them to increase use.

### 1.1 Technology adoption literature

Early models for technology adoption (TA), such as the Technology Acceptance Model (TAM) [20], and its successors TAM2 [21], TAM3 [22], have been used in may disciplines, including health, to explain TA. They consider TA from a behavioural perspective, in an organisational context, and describe TA being dependent on a user's perception of the technology's usefulness, ease of use, and determinants of these. Later models, such as the Unified Theory of Acceptance and Use of Technology (UTAUT) [23] and UTAUT2 [24], consider TA from a combined behavioural, social, psychological and information systems perspective. UTAUT2 considers TA outside of an organisation and highlights that what motivates users inside an organisation is not the same as what motivates them in the home. These models do not specifically consider older users or users with cognitive impairments, or health related technologies. The Senior Technology Acceptance Model (STAM) [25] and Model for the Adoption of Technology by Older Adults (MATOA) [26] do consider older users, but not those with cognitive impairments, and reiterate that a user's perception of usefulness and ease of use are critical for TA. MATOA illustrates that ease of use become more important as users age, and user empowerment and control are important for older users. Healthcare-specific models such as H-TAM [27] illustrate that compatibility of technology with existing lifestyle is important, and this is supported by other research [27, 28] as well as a scoping literature review [29]. Most recently, a model for digital health app adoption by people with cognitive impairments was proposed, which combines Cognitive Load Theory (CLT) [30] and Social Cognitive Theory (SCT) [31]. The model considers technology adoption from the perspectives of (i) how the human brain processes information and how two different parts of a person's memory contribute to cognition, and (ii) how social context and observational processes influence human learning. The model has not been tested yet, however the authors recently added support for relevance of SCT by publishing evaluation results of a solution for people with memory related mild cognitive impairment which combines SCT, spatial augmented reality and deep learning [32].

There are still gaps in our knowledge. Models such as STAM and MATOA may not describe all factors which are significant for TA by older adults with cognitive impairment, and research is continuing to increase understanding of TA by older adults [33] and by older adults with cognitive impairment [32, 34, 35]. A five-year longitudinal study [36] showed that use of everyday technology does become harder for people with MCI, and this has been confirmed by other research [37]. To fully understand TA and ongoing use by older adults with MCI, we must understand what motivates initial adoption, the influence of usage experiences over time, the impact of users' changing physical and cognitive characteristics [38, 39], and other factors which are subject to change over time such as attitude and needs, which have the potential to impact perceived usefulness and ease of use.

### 1.2 Protection motivation theory

The Protection Motivation Theory (PMT) [40], shown in Figure 1, is a psychological model that describes how a person's behaviour is motivated by fear, and how they assess treats and determine coping strategies. Threat appraisal involves assessment of the severity of the threat, the person's perceived vulnerability, and any rewards they may experience from the perceived threat. Coping assessment involves the person's perception of the efficacy of their response, confidence in their ability to successfully execute the response (Self-efficacy), and any costs or negative effects associated with the response. PMT has been used widely in healthcare to describe responses to perceived health risks. It can be used to describe how older adults with MCI assess threats to their life quality from non-use of technology, and how they select preferred coping strategies.

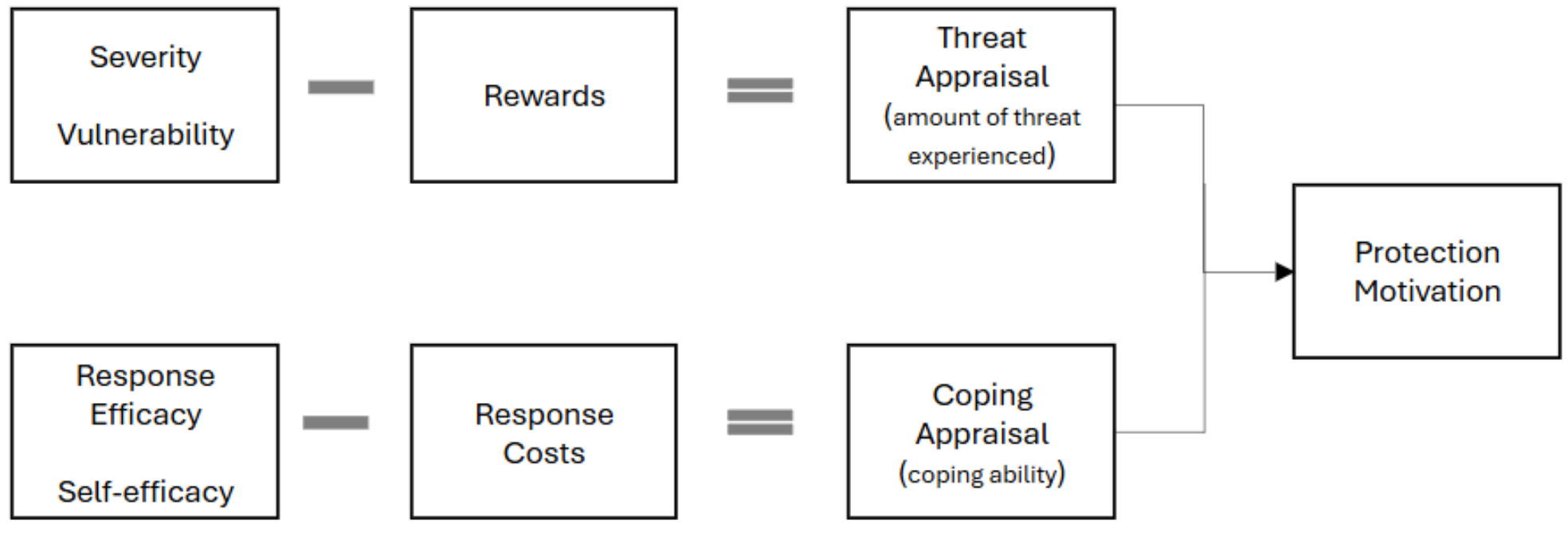

Figure 1: Protection Motivation Theory

### 1.3 Coping strategies used by people with MCI

Coping strategies used by people with MCI include reliance on structure, pre-arranging tasks and activities, maintaining daily routines, use of memory aids, notebooks, memos and lists, and avoiding stress [41-45]. Familiar routines, predictable environments, and structure have been found to reduce anxiety and make older adults with MCI feel calmer, safer, and more in control [46]. Structure is helpful and actively pursued to manage MCI symptoms because it reduces cognitive load, compensates for memory and planning difficulties, and enables the person to maintain their identity, independence and participate in activities they value. Older adults with MCI describe a need for 'normal routine activity patterns' and feel anxious, stressed, and frustrated when emerging time management and organisational difficulties related to MCI disrupt these routines. They feel a stronger sense of disability, which further reinforces their need for predictable, structured days [47, 48].

Technology solutions proposed for older adults with MCI should support their need for structure, routine, predictability, and stability to maintain comfort and reduce stress and anxiety. However, technology is constantly changing, whether through upgrades, enhancements, or new solutions. Each year, a variety of new devices are released onto the market with new features, and innovative solutions and apps are designed, alongside multiple updates to existing operating systems and software. These updates may change the user interface, number of steps or order of steps in a process, functionality offered, security model, and data required to be provided by the user. The changes affect the overall user experience, and users must put in effort to keep up. Changes in technology also affect the outcome of threat assessment related to non-technology use and the coping strategies available for selection. A study involving 2121 mostly-healthy participants found that the younger and middle generations were more likely to be able to keep up with technology changes [49]. Higher education, gender, attitude towards technology, interest, perception of usefulness and usability, and positive experiences increased the likelihood of keeping up. Therefore, targeted strategies are essential to support older adults, and older adults with MCI to keep up with technology. We want to understand the effects of an ever-changing technology landscape on TA and ongoing use for older adults, especially older adults with MCI, who rely on structure, routine, predictability, and stability. We posed the following research questions:

RQ1: How do older adults with MCI and their family members/ carers experience changes to technology?

RQ2: How can we minimise or eliminate negative effects associated with technology change?

This article describes the effects of technology change on older adults with MCI and their family members/carers through the lens of PMT. To our knowledge, this is the first study which explores the impact of technology change on this population.

## 2 MATERIALS AND METHODS

### 2.1 Semi-structured interviews

This qualitative study used semi-structured interviews to explore needs, priorities, lifestyle, technology use, and use experiences of people over 65, with MCI, living in the Australian Capital Territory (ACT). Participants were Australians with MCI, and stakeholders such as family members/carers, and medical and allied health professionals. Participants were sourced using non-random, purposeful, convenience sampling. Recruitment channels were geriatrician clinics, University of Canberra Hospital Memory Assessment Services, independent living villages, the Dementia Australia website, and the Australian Medical Association (ACT Branch) magazine 'Canberra Doctor'. The study required participants with specific characteristics (over 65 with MCI, family member/carer of a person with MCI), was supported by proven theory, and used a comprehensive interview guide, targeted to each stakeholder group. Sample size was determined using 'Information power' [50]. We observed consistency in responses quite early during data-gathering. All participants provided informed consent, and participants with MCI were required to confirm a formal MCI diagnosis by a medical professional. Overall, we interviewed twelve older adults with MCI, six family members/carers, and seven medical and allied health professionals. Participant demographic information is presented in Table 1.

Table 1: Participant demographics and identifiers

| Participant category | ID | Age | Gender | Education | Profession |
|---|---|---|---|---|---|
| Medical/health | 1 | N/A | N/A | N/A | Geriatrician |
| Medical/health | 2 | N/A | N/A | N/A | Nurse/cognitive assessments |
| Medical/health | 3 | N/A | N/A | N/A | Neuropsychologist |
| Medical/health | 4 | N/A | N/A | N/A | Neuropsychologist |
| Medical/health | 5 | N/A | N/A | N/A | Occupational therapist |
| Medical/health | 6 | N/A | N/A | N/A | Speech pathologist |
| Medical/health | 7 | N/A | N/A | N/A | Geriatrician |
| Family/carer | 10 | 80 | Female | Postgraduate | Clinical psychologist |
| Family/carer | 11 | 68 | Female | Year 12 | Public servant |
| Family/carer | 15 | 70 | Female | Postgraduate | Executive management |
| Family/carer | 25 | 70 | Female | Year 10 | Nurse, dental assistant, aged care |
| Family/carer | 27 | 71 | Female | Bachelor's degree | Nurse |
| Family/carer | 28 | 75 | Female | Bachelor's degree | Legal Secretary, practice manager |
| Adult With MCI | 13 | 91 | Male | Year 11 | Army, public service |
| Adult With MCI | 14 | 97 | Male | Postgraduate | Medical doctor, university lecturer |
| Adult With MCI | 16 | 90 | Male | Year 12 | Accountant |

| Participant category | ID | Age | Gender | Education | Profession |
|---|---|---|---|---|---|
| Adult With MCI | 17 | 93 | Male | Bachelor’s degree | University lecturer |
| Adult With MCI | 18 | 73 | Male | Diploma | Accountant |
| Adult With MCI | 19 | 80 | Female | Year 11 | Retail assistant |
| Adult With MCI | 20 | 87 | Female | Year 12 | Nurse |
| Adult With MCI | 21 | 86 | Female | Year 10 | Office work, nurse |
| Adult With MCI | 22 | 79 | Female | Year 12 | Mothercraft nurse |
| Adult With MCI | 23 | 82 | Male | Postgraduate | Barrister, solicitor |
| Adult With MCI | 24 | 74 | Male | Year 11 | Army (20 years), public servant |
| Adult With MCI | 26 | 68 | Male | Bachelor’s degree | Nurse |

Interviews were conducted by the first author, recorded and transcribed verbatim. All interviews were individual, apart from one interview which involved a participant with MCI and their partner to manage language limitations. Interviews were face-to-face and lasted on average 60 minutes. Participants first provided quantitative demographic data (age, gender, marital status, living situation, education, profession, location of birth, languages spoken). The interview guide was designed to examine participants’ needs, priorities, lifestyle, challenges, and experiences with technology.

### 2.2 Analysis of interview data

A ‘bottom-up’, inductive, reflective, thematic analysis strategy was used to organise and code the interview data, as described by [51-53] and guided by [54]. The possible effect of researchers’ prior experiences and biases on outcomes were reduced by continual reflection and review. The analysis was based on constructionist, relativist and experiential motivations, giving regard to how participant’s (i) experiences, identity, meaning and reality are shaped by society and its norms, (ii) truth and values depend on perspective and context, and are not absolute, and (iii) meaning is based on lived experience. Coding of interviews was initially completed by the first author. Each interview was listened to in its entirety at least once to ensure understanding, and then the transcript was tagged with appropriate codes following a second listening. Coding was both semantic and latent, and the first author extracted key quotes reflecting the code sentiment. In parallel, two authors progressively coded and reviewed a random half of the interview transcripts completed by the first author for each participant category and provided feedback on the allocated codes. The first author applied suggested amendments and continually reviewed other interview transcripts and codes to ensure consistency. Codes were generated into sub-themes and themes based on meaning and finalised upon consensus. The first author maintained open dialogue with the other authors throughout the analysis and coding stages.

## 3 RESULTS

Analysis of interview data generated four themes and related sub-themes which are described in [35], and summarised in table 4 in Appendix 1. This article focusses on theme 4 (Technology can empower or be a threat) and sub-theme 4.4 (Frequent technology change creates stress and anxiety, reduces confidence and self-efficacy, and is a threat to quality of life). Codes are used to attribute quotes to participants. Participants with MCI are identified as M1, M2, etc., family members/carers are identified as F1, F2 etc., and medical and allied health professionals are identified as H1, H2, etc. Quotes are shown in italics, and brackets within italics indicate explanatory text for the quote.

Most people with MCI and their family members/carers expressed little interest in technology or keeping up with advances (F10, F25, F28, M13, M14, M16, M17, M20, M21, M22, M23, M24) nevertheless conveyed that technology enabled their participation in activities and access to services.

> 'I'm falling further, and further behind, and my main thing is I'm just not interested.' (F25)
> 'I just can't see the value in it (social media).' (M16)

Family members/carers said technology provides support and enables them to perform their caring role. People with MCI speak about technology more emotively. They associate it with independence and quality of life (M13, M17, M19, M21, M22, M23), and this creates an incentive to learn and use technology even when their interest levels are low.

> '(Technology is) very important for one's wellbeing and freedom.' (M23)
> 'Life would be so much harder …. Technology is there to support me, so I've got to use that technology to my advantage. If I don't, life gets very, very difficult. All those services that I rely upon now are being withdrawn and replaced by technology.' (M23)
> 'I can pull up my calendar, and that's part of my mind, that's part of my memory (mobile phone).' (M24)

Non-use of technology makes them feel vulnerable to loss of independence and autonomy, which is interpreted as a severe threat to their lifestyle. People with MCI said they are motivated by a fear that if they do not use technology and keep up with changes, their quality of life will decrease (M13, M14, M16, M17, M20, M21, M22, M23, M24). Our analysis illustrates that when presented with Threat Appraisal and Coping Appraisal related to non-use of technology, participants' chosen strategy is to learn and use technology.

> 'I'm not in fear of not keeping up. I'm in fear that my quality of my lifestyle will diminish.' (M23)

Both family members/carers and people with MCI are frustrated with continuous technology change and said it is exhausting having to keep up (F15, F25, F28, M14, M19, M20, M22, M23). Medical and health professionals said people with cognitive decline may find it harder to keep up with technology changes, and the fast pace of change leads to a reluctance to transition to new products (H1, H6).

> 'The need to keep up with changes in technology can be overwhelming.' (F25)
> 'I'm happy with the status quo, usually I don't try new things.' (M22)

Family members/carers said changes to technology result in discomfort, confusion and mistakes (F15, F28). Technology change increases Response Costs because effort is required to keep up with changes. It also reduces Response Efficacy as when the number of errors made increases, perception of technology Self-efficacy and IT literacy decreases.

> 'Changes to provider processes may result in mistakes and people get nervous because they're used to technology working in a particular way.' (F15)

People with MCI described a preference for structure, routine, predictability and consistency, and said it makes them feel comfortable and more in control.

> 'I'm happy with the way things are. I don't want new things. I know that's not good, but that's how I am.' (M22)
> 'I like structure in my life.' (M23)

People with MCI describe technology change distressingly, and said it causes confusion, stress, and anxiety and reduces their level of comfort (M14, M16, M20, M22, M23). Changes to their cognition makes it harder to learn new things and retain information, and changes to technology makes solutions unpredictable and harder to use. This feedback was reiterated by medical and health professionals (H1, H4, H6). This illustrates how technology change adds to Response Costs by adding to cognitive difficulty of tasks and reduces Self-efficacy and Response Efficacy by increasing stress and anxiety and decreasing confidence to use technology. In addition, lack of predictability and harder-to-use solutions decrease their perception of IT literacy.

> 'I'm not good at all, I would say. I mean, I can look after my finances and whatever, but I would say that if anything new comes along, I'd have difficulty. I'm wary of putting numbers in or putting things in, in case it's the wrong thing you know, and I mess up whatever's already there'. (M22)
>
> 'If you don't keep up you get left behind and it's getting harder and harder … and it's only going to get worse as you get older - your capacity to be able to absorb knowledge is not as good as it used to be.' (M23)

People with MCI and family members/carers said low interest in technology results in low IT literacy, which results in low usage (F10, F25, F28, M16) and makes it harder to establish daily patterns that develop into habits (F10, F15, F25, F28, M13, M18, M19, M21, M22, M23).

> 'No, I don't want to get involved (with AI). I mean, perhaps I don't like change.' (M20)

They described a connection and circular relationship between interest in technology, IT literacy, and use. Low IT literacy impacts their ability to use technology independently, which increases dependence on others and creates vulnerability (F10, F15, F28, M13, M14, M16, M17, M19, M20, M21, M22). It also means they only use the capabilities of their device they are familiar with and know how to use (F10, F25, F27, F28, M13, M14, M16, M17, M19, M20, M22, M24), and sometimes when they manage to get things to work, they don't know what they did and cannot replicate it (M16). Low IT literacy or non-use of technology may exclude them from activities or provision of services (F11, F15, F27, F28, M13, M16, M22, M23, M24, M26), and exclusion lowers confidence, self-esteem and comfort levels. Technological change discourages use of technology, which prolongs low IT literacy and reduces Response Efficacy and Self-efficacy.

> 'Well, I think that if I don't interact more, I'll be left behind, as I'm finding, when they change things, that I can't do them. So, I would like someone to, you know, show me.' (M22)

All participants said the rapid pace of technological change is a demotivator for learning technology and keeping up with changes, because the high frequency of change makes things obsolete quickly (H1, F15, F25, M14, M16). This lowers perceived IT literacy, Response Efficacy and Self-efficacy. Participants with MCI said low perception of technology self-efficacy and not being able to use technology has a negative effect on their self-esteem and psychological well-being. They said they are not prepared to put up with discomfort and are not likely to persist with technology if they experience problems, lose trust, or if it does not work the way they expect it to. They will give up rather than persevere (M18, M21).

> 'When it works, I feel absolutely overjoyed, but when it doesn't work, I get cross.' (M17)
>
> 'I think of myself as being dumb when it comes to all this …. I've just never gone into sort of all this technical stuff.' (M20)
>
> 'To me, that just seems such an effort. I just didn't like it because I couldn't work it.' (M21)

Overall, our results demonstrate that older adults with MCI and their family members/carers are torn between the benefits and disadvantages of technology. They use technology to support their lifestyle, enable participation in activities, access support and services, and mitigate loss of independence and autonomy. However, technology change creates stress and anxiety, because they feel vulnerable and afraid that their quality of life will decrease if they do not, or cannot, keep up with changes. Decreased perception of IT literacy and technology self-efficacy has a negative effect on self-esteem. In essence, technology solutions simultaneously enhance and complicate their lives.

## 4 DISCUSSION

Older Australians with MCI and their family members/carers expend effort to become IT literate and attempt to keep up with technology changes even though interest levels are low. Study participants with MCI said technology enables them to remain independent, have autonomy, and live well. Their motivation results from a fear that their quality of life will decline without technology. Existing research has reported similar motivation for use of technology [55-58]. Applying PMT to this scenario, Threat Appraisal involves assessment of the impact of loss of independence and quality of life (Severity) and their susceptibility (Vulnerability), and Coping Appraisal involves perception of their IT literacy and their ability to use technology effectively to respond to the threat (Response Efficacy), their confidence in their ability to use technology (Self-efficacy), and any related negative effects (Response Costs). Constant technology change reduces feelings of technology self-efficacy and perception of IT literacy. It lowers confidence to use technology which results in lower use, less sophisticated usage, and obstructs development of usage habits. The overall effect is a lowering of independence and quality of life.

Self-efficacy impacts patterns of thinking and emotional reactions, and people with low self-efficacy may focus on their deficiencies or magnify potential issues, leading to self-doubt, anxiety, and stress. Albert Bandura defined self-efficacy as 'judgments of how well one can execute courses of action required to deal with prospective situations' [59]. Bandura said a person's perceived self-efficacy will determine which coping behaviours they attempt, and for how long they will persist if they experience obstacles or negative outcomes [60]. Confidence in their own ability to succeed greatly affects a person's motivation, effort, and persistence. Our results illustrate that low self-efficacy results in older adults with MCI being less persistent with technology when they experience issues.

If a person experiences positive outcomes, their self-efficacy will increase. People will try to avoid perceived threats if they do not have confidence in their coping skills, and they will only engage confidently when they feel capable of handling the situation. Our results illustrate that older Australians with MCI and family members/carers experience technology change as a disabler for TA and ongoing use. The frequency of technology change means they do not have an opportunity to become fully familiar and comfortable with technology before it is modified, and they must continuously expend effort to keep up to date. Technology change has a negative impact on their Coping Appraisal through a reduced perception of IT literacy, Response Efficacy and Self-efficacy, and increased Response Costs. This relationship is illustrated in Figure 2. Lower Coping Appraisal increases stress, anxiety, and discomfort and reduces confidence in using technology. This answers RQ1.

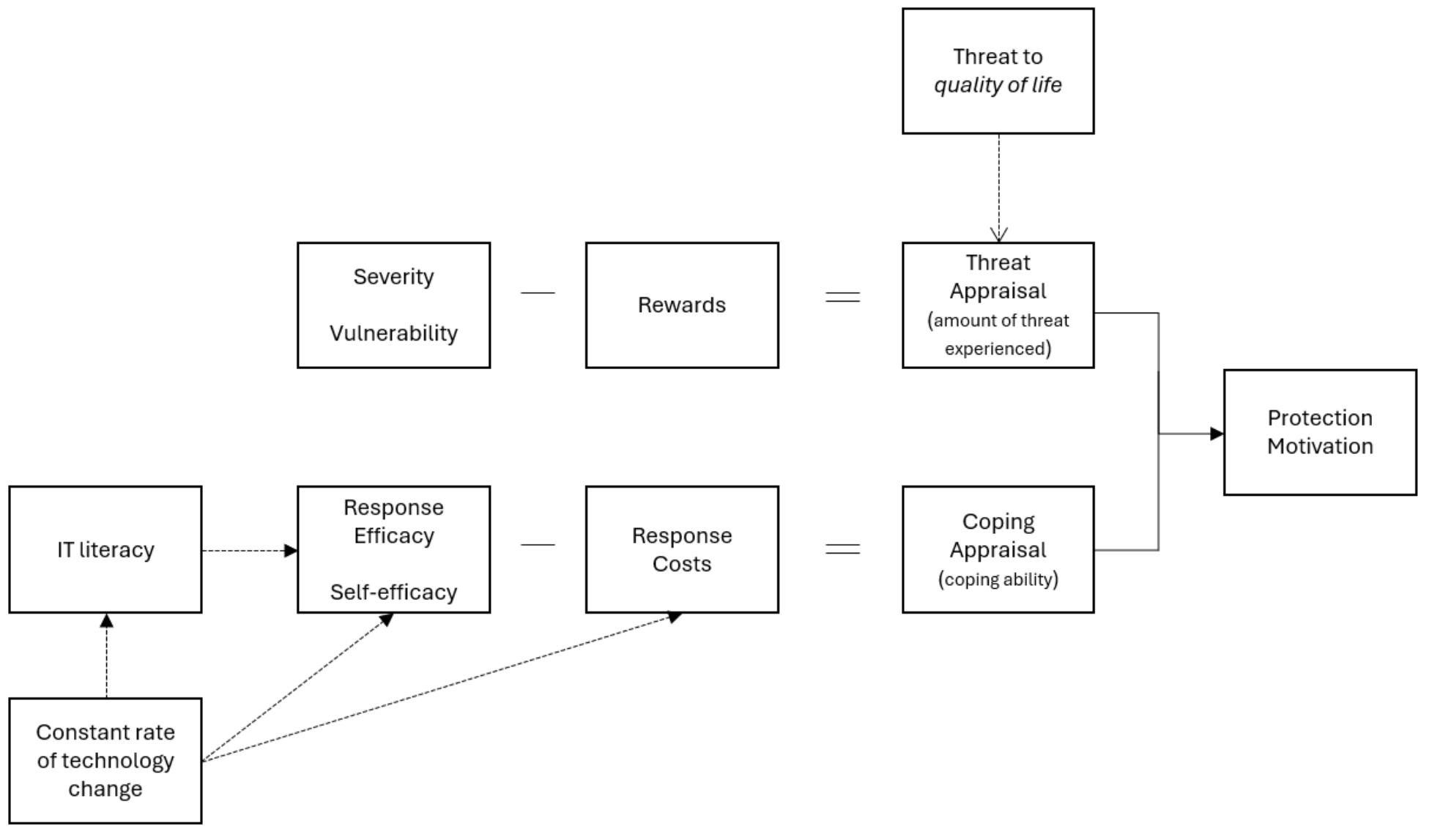

Figure 2: Impact of technology change on perceived IT literacy and Coping Appraisal

Studies of TA have reported perceived self-efficacy is a critical factor for adoption and ongoing use [25-27, 61]. A scoping literature review explained that 30% of included studies identified self-efficacy is important for TA by older adults [62], and other studies note the significance of self-efficacy for sustained use of technology by older adults [63, 64]. In addition, studies of threat assessment describe that Response Efficacy and Self-efficacy have an impact on intention to use technology [65], and that Coping Appraisal is a more powerful predictor of TA intention than Threat Appraisal [66]. Therefore, factors which lower a person's Response Efficacy or perceived Self-efficacy or increase Response Costs will have a more negative effect on TA. Longitudinal support for this is provided by a six month study [67] which found a relationship between technostress and usability, anxiety, and trust for older study participants, which disappeared after familiarity was achieved through extended use. Technostress was found to be a barrier for TA pre-adoption, but not once familiarisation was complete. Constant technology change creates anxiety and reduces users' ability to become familiar with the solution, increase perceptions of ease of use, and establish trust. This has a negative effect on Coping appraisal and allows technostress to continue to act as a barrier for TA.

In addition to a positive effect on TA and ongoing use of technology, existing research suggests that increasing older adults' technology self-efficacy may positively contribute to their self-image and improve their psychological well-being, which contributes to quality of life [68, 69]. A study examining the effect of depression and anxiety on the relationship between self-efficacy and quality of life in people with MCI and dementia concluded that increased self-efficacy may have a positive effect on quality of life for this population by partly reducing depression and anxiety [70]. Proficient and effective use of technology has been shown to increase self-esteem by supporting autonomy and social connection, and raising self-efficacy, which contributes to psychological resilience [71]. Our study outcomes support these conclusions and indicate that increased technology self-efficacy would reduce stress, improve mood and increase comfort levels.

Based on our results, we conclude that technology solutions which maximise users' perceived technology Self-efficacy and IT literacy, will strengthen perceived Response Efficacy, minimise Response Costs, and enhance Coping Appraisal, for older adults with MCI and their family members/carers. This will support their strategy to use technology to reduce threats to independence and quality of life as a result of ageing and MCI. In addition, these solutions will support users to manage changes to technology effectively, minimise any negative impacts to Response Efficacy, Self-efficacy and IT literacy, and positively contribute to TA and ongoing use. We therefore propose the following two design principles for technology solutions used by older adults with MCI and family members/carers to answer RQ2.

Design principle 1: Design to support and maximise feelings of self-efficacy.

Design principle 2: Implement technology change in a manner which minimises or eliminates negative impacts to perceived self-efficacy and IT literacy.

### 4.1 Design principles and guidelines

Bandura described sources of self-efficacy as (i) performance accomplishment, (ii) vicarious experience, (iii) verbal persuasion, and (iv) psychological state, with performance accomplishment having the greatest effect, followed by vicarious experience, verbal persuasion, and psychological state, in that order [60]. Bandura argued that mastering a task to enable successful completion has the greatest positive impact on a person's self-efficacy. Consequently, technology solutions which enable and empower users to achieve successful outcomes will have the maximum positive effect on their perceived Self-efficacy, Response Efficacy and IT literacy. Solutions which minimise or eliminate negative impacts on successful task completion due to technology change will reduce negative impacts on perceived Self-efficacy, Response Efficacy and IT literacy. Based on (i) published research about impacts of MCI, and (ii) a published literature review of technology perspectives and feedback from older adults with MCI [29], we propose design guidelines as listed in Tables 2 and 3 to support and guide implementation of design principles 1 and 2 respectively. Each design guideline is mapped to one of Bandura's four sources of self-efficacy [60]. Given the rapid development of new technologies and the dynamic nature of existing solutions, it is imperative that we can implement technology changes without negatively affecting older users, with or without MCI, whose needs are evolving due to ageing or cognitive decline.

Table 5 in Appendix 2 reframes the design principles and guidelines to align them to coping strategies of adults with MCI.

Table 2: Design principle 1 (design to *support and maximise feelings of self-efficacy*) and related guidelines

| Bandura's source of self-efficacy | Design guideline | Sub-category | Description | References from literature |
|---|---|---|---|---|
| Performance accomplishment (personal experience) | Enable successful experiences | Design methodology for solution creation | Complete user research so needs, preferences, and mental models are understood<br>- Older people with MCI are unique – ageing and MCI do not affect people in the same way<br><br>Follow user-centric design processes and include users in all stages of design activities<br>- Co-design<br>- Consider context and environment of use<br>- Make solution ideas and drafts available to users for evaluation so design decisions are data-driven and evidence-based | [29] |
| | | Familiarity and consistency | Provide personalisation and configuration capabilities to satisfy user preferences and support familiarity (e.g. font size, colour, audio, task progress at own pace, adjust task difficulty) and so solutions are able to adapt over time to changing needs of users due to ageing or cognitive decline<br><br>Ensure solutions are designed for devices favoured by target users, and that preferred interaction modes are available | [38, 72-74] |

| Bandura's source of self-efficacy | Design guideline | Sub-category | Description | References from literature |
|---|---|---|---|---|
| Performance accomplishment (personal experience) | Enable successful experiences | Process design<br><br>Aim for 'Least Advanced, Yet Acceptable' rather than Most Advanced, Yet Acceptable Designs (MAYA Principle) [75] | Ensure processes are simple, short, support a well-defined attainable goal, and update users as they progress<br>- Let users know if there is data or information needed, or if there are time limits, before they start<br>- Do not overcomplicate logins or creation of user accounts<br>- Process steps should match the user's mental model and how they complete tasks<br>- Focus on one task at a time and ensure processes are not too long or have too many steps, so users maintain focus<br>- Allow sufficient time to respond - limit use of time-based responses<br>- Do not design processes with complicated, branching instructions so the user must hold many steps in working memory<br>- Do not assume users know – provide support if needed so users can complete tasks<br><br>Prevent users making mistakes and provide tolerance for error (e.g. intuitive design, supported process steps)<br>- Validate data entry<br>- Use defaults and allow users to update where appropriate<br>- Provide timely warnings to alert and avoid errors<br>- Support users to recover from errors to build resilience and confidence<br>- Confirm actions to provide opportunity to check<br>- Provide ability to undo actions where possible<br>- Provide short, simple, easy to understand, and specific warning and error messages to assist users to avoid or correct errors (e.g. location of issue, why it is an issue and what is needed to address it)<br>- Information text should not disappear too quickly - allow users enough time to read and act<br><br>Provide quick and easy access to content<br>- Prioritise frequently used functions and make these easily accessible<br>- Use plain, familiar and meaningful language so it is understood – do not use jargon or abstract terms<br>- Create navigation that matches the user's mental model and how they categorise content<br>- Keep navigation structures shallow to prevent disorientation<br>- Make all options visible (consider using checkboxes instead of drop-down menus)<br>- Do not display options which are inactive to avoid confusion and working memory load | [75-81] |

| Bandura's source of self-efficacy | Design guideline | Sub-category | Description | References from literature |
|---|---|---|---|---|
| Performance accomplishment (personal experience) | Enable successful experiences | Support cognitive and physical ageing changes | Reduced dexterity and other physical limitations may impact text, touch, gesture or haptic interaction and use of devices<br>- Users may have difficulty using mice, keyboards, trackpads, touchscreens, small screens, interacting with small icons or controls, using devices where pressure or time to push a button is significant<br>- Enable multimodal interaction to provide choice<br>- Consider if it is appropriate to enlarge size of sensitive interaction area<br><br>Progress through process steps is steady and manageable to support changes to executive function and processing speed<br>- Support recognition rather than relying on memory and recall<br>- Guide users through the process steps using short action based tasks and provide indication of progress e.g. visual<br><br>Sensory changes to vision as a result of age or cognitive impairment may affect technology use<br>- Design simple, uncluttered layouts<br>- Do not overload layout with text<br>- Design with considerate use of colour (sharp colours, high contrast, yellow or white background)<br>- Ensure text is easily read, labels are clear, and items stand out<br>- Consider if scrolling is appropriate or if user should see entire screen<br>- Present information to be read left to right, top to bottom<br>- Make controls and widgets large, perceptible, easy to see and interact with<br>- Place interaction elements in centre of screen (to cater for low perception of peripheral elements)<br>- Ensure there is enough space between display items and interactive controls and widgets<br>- Clearly and unambiguously label controls and other interactive items – do not assume icons or symbols are recognised<br><br>Sensory changes to hearing and speech as a result of age or cognitive impairment may affect technology use<br>- Do not rely on sound alone – support modal redundancy<br>- Support subtle differences in speech observed in older adults with MCI (slower speech, smaller chunks, more pauses, longer silences, respond slower to questions, provide shorter answers)<br><br>Create accessible designs<br>- Solutions should not rely on users having particular devices or software (do not create barriers for use)<br>- Enable use of aids such as a stylus, text magnifiers, text readers, headphones if needed or preferred (e.g. add a read aloud function for visually impaired users, enable adaptable colour schemes (e.g. more/less contrasting colours, changing brightness and tone of screen to cater for users with macular degeneration which causes chromatic sensitivity) | [72, 74, 76, 82-84] |

<table>
<tr><th>Bandura's source of self-efficacy</th><th>Design guideline</th><th>Sub-category</th><th>Description</th><th>References from literature</th></tr>
<tr><td rowspan="3">Performance accomplishment (personal experience)</td><td>Enable successful experiences</td><td>Support independence</td><td>Enable configuration and personalisation to support user preferences, flexibility and user empowerment<br><br>Do not make the system controlling or remove choice<br>- Ensure the user remains in charge – doing things by yourself provided satisfaction and creates feelings of empowerment<br>- Do not assume consent without checking<br><br>Do not draw attention to the user's limitations or make them stand out<br><br>Confirm successful completion using affirmative and motivating language<br><br>Ensure access to settings and feedback functions – do not lock users out because you assume they will not understand</td><td>[29, 85-90]</td></tr>
<tr><td rowspan="2">Reduce impact on cognition – processing speed and working memory</td><td>Affordance</td><td>Reduce working-memory demands of perception and categorisation - prototypes and affordances both reduce the amount of information that needs to be held and manipulated in working memory, to support cognition<br>- Affordance Theory (Gibson 1977) describes how people form perceived action possibilities that create fast action-understanding of features<br>- Prototype theory (Rosch 1973) describes how categories are organised around typical exemplars, which allows fast recognition and matching of new inputs to stored long-term 'templates' to speed up recognition<br><br>Processes and tasks where a user can rely on a prototype to quickly categorise an item, and affordances to guide their action, enable users to act based on long-term memory rather than limited, short-term, working memory. Studies with older adults with MCI have shown a reliance for prototype-based, default affordance, and heuristic thinking<br><br>- Buttons, links, controls and other actionable items should be unambiguous, presented consistently, be located where users expect them to be and be recognisable</td><td>[91-94]</td></tr>
<tr><td>Commands and instructions</td><td>Memorising commands is cognitively demanding<br>- Eliminate need to memorise commands, process steps or passwords (support with hints, visual workflow position, wizards)<br>- Avoid use of 'trigger' or 'wake' words</td><td>[95, 96]</td></tr>
</table>

| Bandura's source of self-efficacy | Design guideline | Sub-category | Description | References from literature |
|---|---|---|---|---|
| Performance accomplishment (personal experience) | Reduce impact on cognition – processing speed and working memory | Intuition vs logic | Design low risk tasks to rely on intuition and heuristic thinking to minimise problem solving effort, and only rely on logic and reasoning for high risk tasks<br>- Support high risk, complex and important tasks (e.g. financial) with logic and simple decision steps to prevent errors (e.g. selection of incorrect option due to intuitive thinking)<br>- Consider if use of defaults is appropriate for high-risk tasks<br>- Make all options visible and clear, and make selected choices obvious, to minimise errors<br>- Do not display options which are inactive to avoid confusion and working memory load | [30, 91] |
| | | Minimise interference | Many people with MCI have a poor memory and difficulties with working memory. Interference may have a negative impact on working memory, with interruptions (stimuli that demand attention) having a greater effect compared to distractions (irrelevant stimuli)<br>- Processes should be made up of short process steps which accommodate the user's need for a break<br>- Ensure processes are not too long or have too many steps, so users can maintain focus<br>- Users should be able to save their work and return to it without losing what has already been completed<br>- Designs should prompt users to save work to avoid loss<br>- Visual design should keep users focused and on task, not distract<br>- Avoid loud noises and too many animations or popups because these could distract<br>- Timing of messages and feedback should not interfere with process flow | [97] |
| | | Modal redundancy | Mitigate effects of age-related decline on working memory and aid understanding by communicating information using modal redundancy<br>- Do not rely on visual, sound or haptic interaction only<br>- Provide multisensory and redundant feedback (for example, visual, sound, and tactile feedback) to confirm actions (e.g. vibrate and indicate visually)<br><br>Do not require complex gestures difficult to enact owing to physical ageing (e.g. tremors, shaking) | [98, 99] |
| | | Perception | Stress on perception creates cognitive load (link exists between perception and cognition in old age, in terms of both impact on task performance and age-related decline- degraded input leads to a higher load on cognition, reducing resources available for cognitive processing)<br>- Support perception (vision, hearing, speech, haptic) through considerate design as listed in other parts of this table<br>- Design for multi-sensory inputs and outputs<br>- Use multi-modal interfaces where possible and appropriate | [76, 100] |

| Bandura's source of self-efficacy | Design guideline | Sub-category | Description | References from literature |
|---|---|---|---|---|
| Performance accomplishment (personal experience) | Reduce impact on cognition – processing speed and working memory | Process design | Do not design processes with complicated, branching instructions so the user must hold many steps in working memory<br>- Do not overcomplicate logins or creation of user accounts<br><br>Do not include advertisements or other distractions<br><br>Support recognition rather than relying on memory and recall<br><br>Ageing reduces attentional resources, which makes multi-tasking more difficult and has a negative impact on working memory (this leads to deficits in dual task processing)<br>- Solution designs should consider if multi-tasking is appropriate or if simple, sequential tasks are more suitable (single steps will not distract users and help them to maintain attention and focus)<br><br>Limit the number of items in lists (the number of objects the average person can hold in working memory is about seven)<br>- Complete user research to understand user categorisation schema and slowly narrow down options | [77, 78, 87] |
| | | Support memory and comprehension with images | People with MCI have intact ability to extract and use gist information (conceptual meaning of the item) and pictures support gist-based memory extraction<br>- Use stereoscopic images (realistic images with more detail) rather than flat images<br>- Use simple, easy to recognise visuals and symbology (e.g. + means add) because recognisable images reduce anxiety and make affordances obvious<br>- Include short text descriptions or labels for images to aid recognition if possible and appropriate<br>- Ensure visuals are relatable to target users and not intimidating or scary<br>- Do not use flashing or flickering and ensure any moving visuals (e.g. cartoons) do not move too fast to cause harm | [101-104] |
| | | User interface | Visual complexity of interfaces adds to cognitive load for elderly users<br>- Design simple, uncluttered layouts and do not overload with text to avoid cognitive load and reduce stress<br>- Use high contrast colours, and yellow or white backgrounds<br>- Ensure text is easily read, labels are clear, and items stand out<br>- Clearly and unambiguously label controls and other interactive items – do not assume icons or symbols are recognised<br>- Use consistent navigation<br>- Make controls and widgets large, perceivable, easy to see and interact with<br>- Use multiple modalities to attract attention and communicate cues - provide redundancy of cues (e.g. text, icons, audio)<br>- Allow personalisation of interfaces if possible | [105, 106] |

| Bandura's source of self-efficacy | Design guideline | Sub-category | Description | References from literature |
|---|---|---|---|---|
| Performance accomplishment (personal experience) | Reduce impact on cognition - sensory | Captions and emotion | Cognitive load theory implies any additional linguistic layer should be minimal<br><br>Emotion–cognition and self-reference research with older adults and adults with MCI involving memory limitations suggests short, emotionally supportive captions may aid meaning-making and memory if they replace text | [99, 105, 107-109] |
| | | Language | Use plain, meaningful and easy to understand language to maximise understanding and reduce confusion<br>- Do not use jargon or abstract terms<br>- Clearly and unambiguously label controls and other interactive items – do not assume icons or symbols are recognised | [76, 110-112] |
| | | Speech and listening | Ensure audio or speech interaction does not have an adverse effect on working memory or add to cognitive load - research shows that cognitive effort is needed for spoken work recognition and listening tasks, and words or accents which are not understood add to cognitive load<br>- Reduce listening effort by using short, simple sentences, and familiar, predictable words<br>- Enable options for personalisation (e.g. accent, voice pitch) and configuration (e.g. adjust speed of audio)<br>- Make tempo consistent and ensure there is enough time between utterances so system or user are not interrupted<br>- Do not rely on audio or speech in noisy environments | [72, 74, 76, 82, 113-116] |
| | | Speech, listening and process | Extrinsic cognitive load can impair recognition of spectrally degraded spoken words - solutions should not require spoken sentence comprehension or listening during processes which require high cognitive effort | [115, 117] |
| | Show a clear link between user action and outcome | Do not undermine competence | Support user's actions through the process – do not take over tasks (guide users with step-by-step cues, checklists, previews of consequences)<br>- Confirm actions<br>- Do not assume consent without checking<br>- Provide ability to undo actions if possible<br>- Enable users to navigate ahead or skip steps if appropriate to cater for different skill levels (jump around)<br>- Provide choice - implement personalisation and configuration capabilities for flexibility and user empowerment | [87] |
| | | Make success obvious and visible | Provide immediate or regular feedback, confirmation and success messages<br>- Confirm successful actions and completion of tasks using multisensory and redundant feedback (e.g. vibrate and visual)<br>- Use affirmative and motivating language which is age appropriate | [8, 118-120] |
| Vicarious experience | Show similar others succeeding | Visuals and images in user interface, training and documentation | Use images which include similar others in familiar surroundings and participating in common hobbies or interests<br>- Instructional videos should use the user's demographic for relatability | [103, 121] |

| Bandura's source of self-efficacy | Design guideline | Sub-category | Description | References from literature |
|---|---|---|---|---|
| Verbal persuasion | Encouragement and feedback | Use positive and affirmative language | Use affirmative and motivating language that is age appropriate<br>- Provide short, positive and encouraging messages<br>- Provide immediate or regular feedback at appropriate times in the process<br>- Do not infantilise language or use stigmatising terms<br>- Ensure warnings and errors are informative not frightening, and assist users to avoid or correct errors | [85, 110, 111, 122] |
| Psychological state (reduce the emotional reactions which can lower feelings of self-efficacy) | Prevent user fatigue | Considerate implementation of processes and feedback | Break up processes into simple short steps which have defined goals and accommodate users' need for a break<br>- Enable users to save work completed and return to the task at a later time without losing what has already been done<br>- Prompt users to save work to avoid loss | [80] |
| | | Minimise or eliminate interference and interruptions | Interference has a negative impact on working memory, with interruptions (stimuli that demand attention) having a greater effect compared to distractions (irrelevant stimuli)<br>- Avoid loud noises and too many animations or popups because these could distract<br>- Avoid interference between multimodal outputs (e.g. auditory and written instructions)<br>- Visual design should keep users focused and on task, not distract<br>- Timing of messages and feedback should not interfere with process flow | [97] |
| | Eliminate fear of making mistakes | Considerate implementation of processes, feedback and messages | Prevent users making mistakes and provide tolerance for error (e.g. intuitive design, supported process steps)<br>- Validate data entry<br>- Use defaults where appropriate<br>- Provide timely warnings to alert and avoid errors<br>- Support users to recover from errors to build resilience and confidence<br>- Confirm actions to provide opportunity to check<br>- Provide ability to undo actions where possible<br>- Provide short, simple, easy to understand, and specific warning and error messages to assist users to avoid or correct errors (e.g. location of issue, why it is an issue and what is needed to address it)<br>- Ensure warnings and errors are informative not frightening, and assist users to avoid or correct errors<br>- Text should not disappear too quickly - allow users enough time to read and act | [29] |
| | Eliminate fear of making mistakes | Support and help | Provide quick and easy access to help<br>- Provide context based help<br>- Ensure help and instructions are easily available from every part of solution<br>- Include training material and a glossary if appropriate | [29] |

Table 3: Design principle 2 (implement technology change to *minimise or eliminate negative impacts to perceived self-efficacy and IT literacy*) and related guidelines

| **Bandura's source of self-efficacy** | **Design guideline** | **Sub-category** | **Description** | **References for literature** |
|---|---|---|---|---|
| Performance accomplishment (personal experience) | Enable successful experiences | Familiarity and consistency | Provide personalisation and configuration capabilities so users can retain preferences set in previous version to support familiarity (e.g. font size, audio, task progress at own pace, adjust task difficulty)<br>- Maintain status quo if possible and practical (e.g. keep language, navigation, layout, process steps) | [38, 72-74] |
| | | Support independence | Provide choice<br>- Suggest rather than override (e.g. 'Would you like to add this to your homepage?' rather than automatically adding it)<br>- Implement personalisation and configuration capabilities for flexibility and user empowerment<br>- Provide options to implement change or not if possible<br>- Enable ability to rollback if possible<br>- Do not assume users know – provide support if needed so users can complete tasks | [29, 123] |
| | Reduce impact on cognition | Intuition vs logic | Intuitive design helps people with MCI to learn how to use new solutions because it minimises cognitive load by relying on existing cognitive schemas rather than expending effort to build new ones | [30] |
| | | Minimise learning effort | Avoid need for problem solving because it creates cognitive load and leaves less resources for learning<br>- Use familiar and recognisable visuals and symbology to reduce anxiety and make affordances obvious<br>- Design changes and updates so minimal new learning is required<br><br>Provide overview of changes and highlight location of changes e.g. embed short, contextual walkthroughs after updates (e.g. 'Summary of changes') that appear once automatically but can be accessed at later times | [30] |
| | | Presentation of training information | Communicate information using redundancy across modalities (e.g. audio and text) – this may have greater training impact by mitigating the effects of age-related decline in working memory and reduce cognitive load associated with translation of styles<br>- repeat key information, to reinforce and compensate for attention and memory deficits, using multiple ways<br><br>Include worked examples in training material rather than relying on problem solving<br><br>Provide ability to replay interactive training, if needed<br><br>Provide step-by-step instructions with short messages, to aid comprehension and support attention and working memory | [98, 99] |

| Bandura's source of self-efficacy | Design guideline | Sub-category | Description | References for literature |
|---|---|---|---|---|
| Psychological state (reduce the emotional reactions which can lower feelings of self-efficacy) | Eliminate fear of making mistakes | Communicate changes with sufficient lead time | Provide advance notice of changes with enough detail to inform, but not overwhelm<br>- Offer options to trial changes if possible<br>- Use familiar and recognisable visuals and symbology to reduce anxiety and make affordances obvious | [123] |
| | | Familiarity and consistency | Complete user research to understand existing user processes and priorities to be able to meet needs and integrate new solutions into established routines<br>- A positive user experience is an enabler for technology acceptance | [29, 35] |
| | | Support and help | Provide quick and easy access to help<br>- Provide context based help<br>- Ensure help and instructions are easily available from every part of solution | [29] |

## 5 FUTURE WORK

Future work should rigorously test the proposed design principles and guidelines by using them to build several technology solutions for older adults with MCI to evaluate their effectiveness and use evaluation feedback to iterate and refine solutions. We should also explore how changing the solution purpose or technology practically impacts guidelines, and how design guidelines may differ for different populations or contexts of use. It would also be useful to consider different categories of technology changes to investigate if they all have the same impact on older adults' perceived self-efficacy and IT literacy, and if the method and channel used to communicate change has a significant impact.

## 6 STRENGTHS AND LIMITATIONS

Strengths of the study include (1) being user-focussed, (2) study participant characteristics being specific and well defined, and (3) interview guides being targeted per stakeholder role. The design principles and related guidelines consider the user's overall experience rather than only the solutions' usefulness and ease of use. They consider the impact of technology use on the user's emotional and psychological state, and aim to improve self-efficacy, confidence and self-esteem. In addition, they are applicable to all older users, not only those with cognitive impairments, and recommendations are practical, actionable and based on proven theory and research.

Limitations of the study include that it involved a relatively small number of participants with MCI and their family members/carers from a specific Australian locale. Each person was interviewed only once and only involved English-speaking participants. It would be beneficial to capture data from a broader demographic including other parts of the world, to validate our conclusions as well as cultural differences. The proposed design principles and guidelines are theoretical and have not been validated, however they are grounded in established TA and self-efficacy theory, and supported by studies with older participants with MCI, as well as our own research outcomes.

## 7 CONCLUSION

Supporting older adults with MCI to have autonomy, and live independently with quality of life, has benefits not only for the people with MCI, but also to their family members/carers, to society, and to the health care sector in general. Analysis outcomes of semi-structured interviews with older Australians with MCI, their family members/carers, and medical and allied health professionals indicate interest in technology is low, and adoption and ongoing use is motivated by need and a desire for independence and life quality. Changes to technology are too frequent, and cause frustration, anxiety, and stress. Frequent technology change lowers perceived IT literacy, technology self-efficacy, and confidence to use technology. PMT can be used to illustrate how technology change negatively affects Coping Appraisal when technology is selected as a strategy to manage threats to independence and quality of life. Changes to technology increase Response Cost, and decrease perception of Self-efficacy, Response Efficacy, and IT literacy. Technology solutions which maximise a user's perceived self-efficacy and IT literacy and minimise or eliminate the negative impacts of technology change, will positively impact TA and ongoing use for older adults with MCI and their family members/carers. Two design principles and related design guidelines are proposed, aligned with Bandura's theory of self-efficacy and supported by existing studies and literature. Future work should incorporate these guidelines into technology solutions designed for older adults with MCI so they can be evaluated with target users, refined and expanded.

### Statement of ethics

The research has ethics approval from <removed for review process>.

### Conflict of interest statement

The authors declare no conflict of interest.

### Data availability

Participant identifying data is not publicly available due to privacy and ethical considerations. Complete results of interview analysis is available by request.

## APPENDIX 1 – THEMES AND SUB-THEMES RESULTING FROM INDUCTIVE, REFLECTIVE, THEMATIC ANALYSIS OF INTERVIEW DATA

Table 4: Themes and sub-themes arising from thematic analysis of interview data

| Theme | Sub-theme |
|---|---|
| 1. People are unique and have individual needs based on context | 1.1 The goal is independence, autonomy and quality of life<br>1.2 Needs are individual, specific and multi-faceted<br>1.3 Mixed feelings about technology – it yields rewards but has costs |
| 2. People use a variety of tools and personalised strategies to support independence | 2.1 Simple non-technology based strategies are often effective<br>2.2 Technology based strategies must be appropriate and satisfy performance and effort expectancy<br>2.3 Frequency of use develops habits<br>2.4 First impressions, IT literacy and self-efficacy affect ongoing use |
| 3. Selection and use of technology depends on social influence, availability of support, tasks, context of use, and own preferences | 3.1 Family and significant others influence technology selection because they are relied upon for help and support<br>3.2 Biophysical ageing restrictions, context of use, and task, determine device and interaction preferences |
| 4. Technology can empower or be a threat | 4.1 Technology can enrich and support independence<br>4.2 Low technology knowledge, experience and IT literacy may exclude<br>4.3 Low technology knowledge, experience and IT literacy makes people feel vulnerable or unvalued<br>4.4 Frequent technology change creates stress and anxiety, reduces confidence and self-efficacy, and is a threat to quality of life |

## APPENDIX 2 – THEMES AND SUB-THEMES RESULTING FROM INDUCTIVE, REFLECTIVE, THEMATIC ANALYSIS OF INTERVIEW DATA

Table 5 describes design principles and guidelines to support coping strategies of adults with MCI.

| Coping strategy | | Guideline | Notes |
|---|---|---|---|
| Structure (reduce need to make decisions) - stability AND familiarity, regularity, predictability | General | Use industry wide/common design and interaction patterns to support predictability and familiarity with other technology which may be in use | |
| Structure (reduce need to make decisions) - stability AND familiarity, regularity, predictability | Change | Consider technology solutions wholistically during initial design and implementation planning - releasing solutions in stages or adding to available capabilities may cause stress and anxiety for older adults with cognitive impairment<br>- technology change is a major disabler for adoption and ongoing use because it creates stress and confusion, lowers perception of technology self-efficacy and literacy, and reduces motivation to learn<br>- staged rollouts may not be the best approach (these are often employed by government agencies (when they release the 'Minimal Viable Product (MVP)' or organisations who want to get products to market early)<br><br>This is an important factor to consider for government services when they are creating new technology-based services or updating existing services with new capabilities, changing processes, updating terminology, etc. | Older adults say technology change causes frustration and stress and it is exhausting to keep up with changes<br><br>Older adults with MCI describe technology change alarmingly and say it causes confusion and anxiety because it makes solutions inconsistent, unpredictable, and harder to use, and weakens their coping strategies to manage MCI symptoms. It creates fear for their independence and QoL.<br>- cognitive changes related to MCI make it harder to learn new things, and technology change degrades their perception of IT literacy and technology self-efficacy<br>- medical/health professionals support this, and add that adults with cognitive decline find it harder to adapt and keep up with technology changes<br><br>Both groups say technology change leads to a reluctance to transition to new products, discourages use of technology and may prolong low IT literacy |
| Structure (reduce need to make decisions) - stability AND familiarity, regularity, predictability | Content | Older adults with MCI may experience impacts to vision and difficulty 'seeing the big picture' if some information is not available<br>- consider if any content needs to be visible at all times to reinforce structural expectations and also reduce cognitive load<br>- provide quick and easy access to important/frequently used functions and information e.g. place in a prominent location/top of screen and make them easily accessible | |
| Structure (reduce need to make decisions) - stability AND familiarity, regularity, predictability | Content - language | Use simple, meaningful and non-threatening language consistently (content, labels, headings, messages, etc.)<br>- plain, familiar and meaningful language is more easily understood<br>- do not use jargon or abstract terms and explain all abbreviations<br><br>Use of simple and meaningful language consistently, also has a role in reducing cognitive load | |
| Structure (reduce need to make decisions) - stability AND familiarity, regularity, predictability | Content - layout | Present content using a consistent and clear hierarchy of headings and labels<br>- provide quick and easy access to important/frequently used functions and information e.g. place in a prominent location/top of screen and make them easily accessible<br>- do not overload layout with text - this may overwhelm and cause stress | |

| Coping strategy | | Guideline | Notes |
|---|---|---|---|
| Structure (reduce need to make decisions) - stability AND familiarity, regularity, predictability | Lifestyle, routines, habits and workflow | Consider how the solution will fit into users' existing lifestyles and routines (take a user centric approach to design e.g. who is the target user, what technology/devices do they use, what are expected scenarios of use, how will the users interact with the solution (touch, text, voice, etc.), where will they use the solution, etc.)<br>- routines can build a connection between a person's cognitive constraints and their use of technology<br>- consider how amendments can be made to support existing structure or routines rather than disrupt by requiring change<br><br>Habits are actions that are automatically prompted by contextual cues associated with their previous performance. Once the action becomes tied to an external cue, it requires less conscious attention and motivation. This makes **habits cognitively efficient**, because automation of actions frees mental resources for other tasks. As a result, habits may persist even after deliberate motivation or interest has vanished. | Many well-established, automatic routines are preserved in the early stages of MCI because they rely more on procedural memory ('how to do things') than on episodic memory ('remembering events')<br><br>MCI can disrupt established habits or make them less reliable when consistency is reduced, circumstances change, a relied-on cue is removed, planning or organisation is impaired, task switching becomes more difficult, or the person experiences stress<br><br>MCI may affect memory, language, executive functioning, visuospatial ability, flexibility or ability to adapt. Therefore, established habits may be better preserved than those needing adaption. Long-standing routines are often maintained, whereas habits that rely on learning new information or adjusting to change are more likely to be disrupted |
| Structure (reduce need to make decisions) - stability AND familiarity, regularity, predictability | UI - multiple items | Older adults with MCI may experience impacts to vision<br>- difficulty seeing more than one thing at a time<br>- difficulty being able to pick something out from a cluster of items | |
| Structure (reduce need to make decisions) - stability AND familiarity, regularity, predictability | UI - colour | Older adults with MCI may see colours less sharply or they may have less contrast sensitivity<br>- eliminate visual clutter - use white space effectively<br>- ensure important items stand out<br>- use sharp colours, high contrast, yellow or white background<br>- use colour consistently<br>- see also items related to 'Sensory - vision' | |
| Structure (reduce need to make decisions) - stability AND familiarity, regularity, predictability | UI - widgets and controls | Older adults with MCI may have low perception of peripheral elements<br>- make widgets and controls large, perceptible, and easy to see and interact with<br>- place main interaction elements in centre of screen<br>- ensure there is enough space between items and interactive elements<br>- consider making the 'active' area for the widget or control bigger to support users' ageing limitations (also, some people want to see the label so they stay clear of the words)<br><br>Label widgets, controls, and other interactive elements consistently, unambiguously and use language that is simple and meaningful | |
| Reduce cognitive load | Content | Reduce cognitive effort associated with reading (see also 'Sensory' and 'UI')<br>- present content in small, easily manageable 'chunks' so as not to overload cognition<br>- use short phrases<br>- use clear and consistent layouts | |
| Reduce cognitive load | Interference and distraction | Older adults with MCI may have difficulty blocking out noise or concentrating on a voice - distractions reduce focus and create cognitive load<br><br>Minimise interference or distraction to reduce negative impacts on working memory<br>- consider timing of messages within a workflow to maintain focus and avoid distractions<br>- avoid loud noises and too many animations because these distract<br>- avoid large or colourful graphics<br><br>Avoid interference between multimodal outputs (e.g. auditory and written instructions) | |

| Coping strategy | | Guideline | Notes |
|---|---|---|---|
| Reduce cognitive load | Intuition rather than logic | Avoid need to problem solve because problem solving creates cognitive load and leaves less resources for logic or learning<br>- intuition reduces reliance on processing speed and working memory<br><br>Rely on intuition rather than logic for simple tasks but more important or complex tasks may need to rely more on logic to avoid mistakes<br>- design low risk tasks to rely on intuition and heuristic thinking to minimise problem solving effort, and only rely on logic and reasoning for high risk tasks<br>- support high risk, complex and important tasks (e.g. financial) with logic, simple decision steps, confirm progress, and check inputs to prevent errors | Intuitive design helps people with MCI use new solutions with minimal problem solving - it minimises cognitive load by relying on existing cognitive schemas rather than expending effort to build new ones |
| Reduce cognitive load | Lifestyle and workflow | Established habits are cognitively efficient, because automation of actions frees mental resources for other tasks<br>- consider how the solution can be mapped as seamlessly as possible onto an existing, ingrained habit or daily routine or ritual to avoid cognitive friction<br>- if a solution anchors itself to an existing anchor, the procedural memory of the existing routine reduces the cognitive effort needed to remember (minimising cognitive load) | Habits are actions that are automatically prompted by contextual cues associated with their previous performance. Once the action becomes tied to an external cue, it requires less conscious attention and motivation. This makes habits cognitively efficient, because automation of actions frees mental resources for other tasks. As a result, habits may persist even after deliberate motivation or interest has vanished. |
| Reduce cognitive load | Lists, selection, options | Limit the number of items in lists to avoid overwhelming users<br><br>Consider the order that selections are presented because older adults with cognitive impairment may not remember all items in the list and select the first acceptable option<br><br>Make all options visible because hidden options may confuse (e.g. consider using checkboxes instead of drop-down lists, etc.)<br><br>Avoid the use of multiple levels to expose data or information<br><br>Do not display options which are inactive to avoid confusion and working memory load | For individuals with MCI, working memory may be both a core area of impairment and the mechanism through which cognitive load is experienced (working memory is sometimes called short-term memory)<br><br>Reduced cognitive capacity leads to increased reliance on heuristics and satisficing which leads to greater likelihood of choosing highly visible, early, or default options |
| Reduce cognitive load | Navigation | Older adults with MCI may lack abilities to orient in websites or navigate around<br>- orient user in the structure and make it obvious where they are (e.g. visually, using breadcrumbs, step numbering (e.g. 3 of 5), etc.)<br><br>Ensure navigation is visible and recognisable and ensure it matches the user's mental model<br>- support how users categorise content<br>- ensure navigation is consistent<br>- avoid the use of multiple navigation levels - keep navigation structures **shallow or linear** to prevent disorientation (which may cause stress)<br>- ensure users are able to exit or return to starting position (so they do not feel 'stuck') | |
| Reduce cognitive load | Process - multiple platforms or mediums | When designing an end-to-end process requiring use of multiple platforms or mediums over time<br>- use one primary entry point and provide a clear overview of the process and steps<br>- ensure user knows what step they are up to and tell them what is next, timing, etc.<br>- avoid the need to memorise logins, URLs, phone numbers, etc.<br>- use consistent layouts and design patterns as much as possible<br>- create a connection between the platforms to ensure context is understood e.g. use the same language, colour scheme, visuals, perhaps use a visual/graphic to identify the service or process, etc.<br>- build in available human support in case it is needed<br><br>Consistency and predictability are very important | |

| Coping strategy | | Guideline | Notes |
|---|---|---|---|
| Reduce cognitive load | Process | Break up complex tasks into multiple steps or tasks to help compensate for limited working-memory resources<br>- make processes simple, short, with discrete steps, and support well-defined attainable goals<br>- avoid branching instructions that require the user to hold many steps in working memory<br>- avoid implementing processes or steps with time limits<br>- allow extra time for processing<br>- ensure progress through process steps is steady and manageable to support changes to executive function and processing speed<br>- provide users with control over the pace and how they progress through the process<br>- allow users to save work if appropriate and prompt to save to avoid loss<br><br>Aim for 'Least Advanced, Yet Acceptable' rather than Most Advanced, Yet Acceptable Designs (MAYA Principle) | |
| Reduce cognitive load | Process | Focus on one task at a time to help users maintain focus<br>- avoid multi-tasking if simple, sequential tasks are (single steps will not distract users and help them to maintain attention and focus)<br>- avoid task switching<br><br>Memory demands create cognitive load | Ageing reduces attentional resources, which makes multi-tasking more difficult and has a negative impact on working memory (this leads to deficits in dual task processing) |
| Reduce cognitive load | Process | Provide progress updates to keep users motivated<br>- guide users through the process steps using short action based tasks and provide indication of progress e.g. visual | |
| Reduce cognitive load | Process | Support how users actually behave so they do not have to problem solve or change how they complete tasks<br>- process steps should match the user's mental model | |
| Reduce cognitive load | Process | Use multiple modalities to attract attention and communicate cues<br>- provide redundancy of cues (e.g. text, icons, audio) | |
| Reduce cognitive load | Sensory | Stress on perception creates cognitive load (there is a link between perception and cognition in old age, in terms of both impact on task performance and age-related decline<br><br>Degraded sensory input leads to a higher load on cognition and reduces resources available for cognitive processing)<br>- Reduce stress related to vision or to auditory, touch or text interaction (see 'Sensory-audi and speech' and 'Sensory - vision')<br>- Support multi-modal options where possible and appropriate | Older adults with MCI may experience impacts to perception<br><br>Vision<br>- Less sharp<br>- Less contrast sensitivity<br>- Difficulty seeing more than one thing at a time<br>- Difficulty seeing the big picture if all required information is not displayed/available<br>- Difficulty being able to pick something out from a cluster of items<br><br>Reading<br>- Surface Dyslexia. People with dementia can experience surface dyslexia, meaning it can be difficult to read irregular words and classify stimuli as words or not words<br>- Some people can understand written information better than spoken information<br><br>Difficulty blocking out noise and concentrating on a voice<br>Speech and language patterns can change<br>Dexterity and fine motor skills decrease<br>Ability to move body and balance may be affected as MCI progresses |

| Coping strategy | | Guideline | Notes |
|---|---|---|---|
| Reduce cognitive load | Sensory - audio and speech | Cognitive effort is needed for spoken work recognition and for listening tasks<br>- words or accents which are not understood add to cognitive load<br>- reduce listening effort by using short, simple sentences, and familiar, predictable words<br>- enable options for personalisation (e.g. accent, voice pitch) and configuration (e.g. adjust speed of audio)<br>- ensure speech is not too fast<br>- make tempo consistent and ensure there is enough time between utterances so system or user are not interrupted<br>- do not rely on audio or speech in noisy environments<br><br>Extrinsic cognitive load can impair recognition of spectrally degraded spoken words - solutions should not require spoken sentence comprehension or listening during processes which require high cognitive effort | Use of audio and text support can benefit all users, but speech is linear so unimodal speech interaction may cause problems.<br>- audio interfaces may create a new usability issue because they create cognitive load (user is forced to rely on memory and mental agility) - users avoid a keyboard and have to read less text, but they have to comprehend and retain information while making interaction decisions |
| Reduce cognitive load | Sensory - audio and speech | Older adults with MCI may experience changes to their speech and language patterns<br>- speed of speech and volume<br>- response to questions may be slower<br><br>Older adults with MCI may have difficulty blocking out noise and concentrating on a voice<br>- avoid loud noises because these distract and create cognitive load | |
| Reduce cognitive load | Sensory - vision | Present information to be read left to right, top to bottom and in a single column<br>- it reduces sensory load on vision (scanning) and need for decision-making | Older adults with MCI may experience impacts to vision<br>- Less sharp<br>- Less contrast sensitivity<br>- Difficulty seeing more than one thing at a time<br>- Difficulty seeing the big picture if all required information is not displayed/available<br>- Difficulty being able to pick something out from a cluster of items |
| Reduce cognitive load | UI - scrolling | MCI may affect working memory and attention and make it harder to integrate information spread over a long page<br>- avoid excessive scrolling - this may create cognitive load or cause stress<br>- consider if scrolling is appropriate or if user should see entire screen (see "Content')<br>- long scrolling pages may cause users to lose focus, lose their place, or forget information they read earlier | |
| Reduce cognitive load | UI | Visual complexity of interfaces adds to cognitive load for elderly users<br>- design simple, uncluttered layouts and use white space effectively<br>- do not overload with text to reduce stress<br>- use high contrast colours, and yellow or white backgrounds<br>- ensure text is easily read, labels are clear, and important items stand out<br><br>Dense screens showing everything at once can be overwhelming because too much information increases cognitive load<br>- simple layouts are recommended<br>- progressive disclosure may not always be appropriate | There is another side to cognitive load - oversimplifying (for example, by greatly reducing the number of elements in an interface) also causes UX friction because it forces users to think harder since more functionality is hidden and/or assumed.<br><br>Cognitive load can increase with an increase in the number of apps, widgets, and icons and also with their decrease (past a certain point). Complexity as well as over-simplification increases the degree to which users must think about an action or interface |

| Coping strategy | | Guideline | Notes |
| --- | --- | --- | --- |
| Reduce cognitive load | UI - images | People with MCI have intact ability to extract and use gist information (conceptual meaning of the item) and pictures support gist-based memory extraction<br>- use stereoscopic images (realistic images with more detail) rather than flat images<br>- use simple, easy to recognise visuals and symbology (e.g. + means add) because recognisable images reduce anxiety and make affordances obvious<br>- include short text descriptions or labels for images to aid recognition if appropriate but be careful not to add to cognitive load by requiring reading and image recognition at the same time<br>- ensure visuals are relatable to target users and not intimidating or scary<br>- do not use flashing or flickering and ensure any moving visuals (e.g. cartoons) do not move too fast to cause harm | |
| Reduce cognitive load | Tables or lists | Lists and tables may not be the best way to present information to older adults with cognitive impairments - think about other ways to present the information<br>- keep tables and lists as simple as possible<br>- simple lists with pictograms are better than complex tables or matrices<br>- avoid dense grids<br>- use redundancy of information presentation (present the same information multiple ways, don't present some information in one medium and other information in another - this will confuse)<br>- present core information and provide users with a choice if they want to know more - disclose additional information if the user is interested (using 'Find out more' or something similar) | |
| Reduce cognitive load | Videos | Videos should be short (30 - 90 secs) and communicate one idea<br>- narration should be slow and steady (2 words per second) and include pauses<br>- consider the speech and accent used by the narrator because listening effort can create cognitive load | |
| Memory support | General | Do not rely on memory - support recognition rather than relying on memory and recall | |
| Memory support | Lifestyle and workflow | Use existing user routines and schedules as triggers or cues (time, placement/location, activities) | |
| Memory support | Messages | Use redundancy of messaging to support memory limitations<br><br>Messages and information text should not disappear too quickly - allow users enough time to read and act<br><br>Provide real-time feedback for user motivation | |
| Memory support | Process | Eliminate need to memorise data, commands, steps or passwords (adults with MCI often struggle to learn new sequential steps due to decline in short-term memory)<br>- use prompting where appropriate<br>- prefill known data or use defaults to avoid effort and reduce need to remember information<br>- avoid using trigger words or 'wake' words<br>- do not require logins or user accounts unless necessary<br>- consider biometric access where appropriate and possible | |
| Memory support | Process | Enable point in time recording (e.g. create a note at the time a thought occurs) to support memory limitations | |

| Coping strategy | | Guideline | Notes |
|---|---|---|---|
| Memory support | UI - affordances | Use recognisable symbols to make affordances obvious (for widgets, icons, etc.)<br>- use simple, easy to recognise visuals and symbology (e.g. + means add) because recognisable images reduce anxiety and make affordances obvious<br>- include short text descriptions or labels for images to aid recognition if possible and appropriate<br>– do not assume icons or symbols are recognised | Reduce working memory demands of perception and categorisation - prototypes and affordances both reduce the amount of information that needs to be held and manipulated in working memory, to support cognition<br>- Affordance Theory (Gibson 1977) describes how people form perceived action possibilities that create fast action-understanding of features<br>- Prototype theory (Rosch 1973) describes how categories are organised around typical exemplars, which allows fast recognition and matching of new inputs to stored long term 'templates' to speed up recognition |
| Stress avoidance | General | Inform users if any information or documents are needed prior to start of task to avoid stress or loss of work | |
| Stress avoidance | Error prevention and tolerance | Eliminate fear of making mistakes<br>- provide context based help<br>- provide timely warnings to alert and avoid errors<br>- display useful messages using non-threatening language<br>- indicate what valid data 'looks like' for data entry tasks (e.g. use date picker, display example in cell, etc.)<br>- use defaults where appropriate and allow users to update where appropriate<br>- validate data at the point of entry<br>- provide ability to undo or go back to a previous step in the process where possible<br>- support users to recover from errors to build resilience and confidence (ability to correct or undo)<br>- allow for error correction, ensure location of error is made clear and that error messages include information about how to correct the error<br>- information text should not disappear too quickly - allow users enough time to read and act<br>- confirm actions and successful completion of tasks at time of completion<br>- allow users to save work if appropriate and prompt to save to avoid loss | |
| Stress avoidance | Independence and autonomy | Empower users and support self-esteem<br>- doing things by yourself provides satisfaction and enhances feelings of efficacy<br>- do not create designs which draw attention to the user's limitations or make them stand out<br>- confirm actions and successful completion of tasks at time of completion using affirmative and motivating language | |
| Stress avoidance | Process | Allow sufficient time to complete steps and tasks to avoid stress - older adults with MCI generally take longer to complete tasks than older adults without MCI<br>- avoid implementing processes or steps with time limits<br>- allow extra time for processing<br>- provide users with control over the pace and how they progress through the process<br>- messages and information text should not disappear too quickly - allow users enough time to read and act | |
| Support physical ageing restrictions including vision, hearing, manual dexterity | Gesture | Do not require complex gestures or minute, specific touch interactions or text entry on small screens - these may be difficult due to limitations associted with MCI or physical ageing (e.g. arthritis, tremors, shaking) | |
| Support physical ageing restrictions including vision, hearing, manual dexterity | Independence and autonomy | Do not undermine competence - support user's decisions and actions throughout the process<br>- do not take over tasks (guide users with step-by-step cues, checklists, previews of consequences)<br>- do not assume consent without checking<br>- provide ability to undo actions if possible<br>- enable users to navigate ahead or skip steps if appropriate to cater for different skill levels (jump around) | |

| Coping strategy | | Guideline | Notes |
|---|---|---|---|
| Support physical ageing restrictions including vision, hearing, manual dexterity | Independence and autonomy | Implement accessible design to eliminate barriers for use<br>- do not rely on users having particular devices or software<br>- support accessibility tools such as a stylus, text magnifiers, text readers, etc.<br><br>Provide configuration and personalisation options (e.g. text size, colour, etc.) to support user preferences, flexibility and user empowerment<br>- support choice (ensure the user remains in charge )<br>- enable adaptable colour schemes (e.g. more/less contrasting colours, changing brightness and tone of screen to cater for users with macular degeneration which causes chromatic sensitivity)<br><br>Support multimodal interaction options (e.g. text, touch, voice, biometric (fingerprint, face)) to support choice | |
| Support physical ageing restrictions including vision, hearing, manual dexterity | Interaction | Reduced dexterity and other physical limitations may impact text, touch, gesture or haptic interaction and use of devices<br>- users may have difficulty using mice, keyboards, trackpads, touchscreens, small screens, interacting with small icons or controls, using devices where pressure or time to push a button is significant<br>- provide multimodal interaction options (e.g. text, touch, voice, biometric (fingerprint, face)) to support choice<br>- consider if it is appropriate to enlarge size of sensitive interaction area | Dexterity and fine motor skills decrease as people age<br><br>Ability to move body and balance may be affected as MCI progresses |
| Support physical ageing restrictions including vision, hearing, manual dexterity | Sensory - audio and speech | Sensory changes to hearing and speech as a result of age or cognitive impairment may affect technology use<br>- do not rely on sound alone – support modal redundancy<br>- support subtle differences in speech observed in older adults with MCI (slower speech (2 words per second and add 2s of silence in between), short sentences, smaller chunks, more pauses, longer silences, respond slower to questions, provide shorter answers) | |
| Support physical ageing restrictions including vision, hearing, manual dexterity | Sensory - vision | Sensory changes to vision as a result of age or cognitive impairment may affect technology use<br>- design simple, uncluttered layouts<br>- do not overload layout with text<br>- ensure text is easily read and labels are clear<br>- ensure priority content and interactive elements stand out and place in a prominent location/top of screen<br>- use sharp colours, high contrast, yellow or white background<br>- present information to be read left to right, top to bottom<br>- make controls and widgets large, perceptible, easy to see and interact with<br>- place interaction elements in centre of screen to support users with low perception of peripheral elements<br>- ensure there is enough space between display items and interactive controls and widgets so users can select these with accuracy | Vision<br>- Less sharp<br>- Less contrast sensitivity<br>- Difficulty seeing more than one thing at a time<br>- Difficulty seeing the big picture if all required information is not displayed/available<br>- Difficulty being able to pick something out from a cluster of items<br><br>Reading<br>- Surface Dyslexia. People with dementia can experience surface dyslexia, meaning it can be difficult to read irregular words and classify stimuli as words or not words<br>- Some people can understand written information better than spoken information |
| Provide support for carers | General | Provide multi-platform support to aid integration with other technologies in use by user and support network | |
| Provide support for carers | General | Provide multi-user solutions, data sharing options, and real-time updates | |

Table 5: Design principles and guidelines to support coping strategies of adults with MCI